\documentclass[
superscriptaddress,
twocolumn,
 amsmath,amssymb,
 aps,
prl,
floatfix,
a4paper
]{revtex4-2}
\usepackage{graphicx,color}
\usepackage{dcolumn}
\usepackage{bm}
\usepackage{physics}

\usepackage{natbib}
\usepackage{here}
\begin{document}
\title{Large Diamagnetism and Electromagnetic Duality in Two-dimensional Dirac Electron System}
\author{S.~Fujiyama}
\email[Electronic address: ]{fujiyama@riken.jp}
\affiliation{RIKEN, Condensed Molecular Materials Laboratory, Wako 351-0198, Japan}
\author{H.~Maebashi}
\email[Electronic address: ]{maebashi@hosi.phys.s.u-tokyo.ac.jp}
\affiliation{Department of Physics, University of Tokyo, Tokyo 113-0033, Japan}
\author{N.~Tajima}
\affiliation{Department of Physics, Toho University, Funabashi 274-8510, Japan}
\author{T.~Tsumuraya}
\affiliation{POIE, Kumamoto University, Kumamoto 860-8555, Japan}
\author{H-B.~Cui}
\affiliation{RIKEN, Condensed Molecular Materials Laboratory, Wako 351-0198, Japan}
\author{M.~Ogata}
\affiliation{Department of Physics, University of Tokyo, Tokyo 113-0033, Japan}
\affiliation{Trans-scale Quantum Science Institute, University of Tokyo, Bunkyo-ku, Tokyo 113-0033, Japan}
\author{R.~Kato}
\affiliation{RIKEN, Condensed Molecular Materials Laboratory, Wako 351-0198, Japan}
\date{5 May, 2021}
\begin{abstract}

A Dirac electron system in solids mimics relativistic quantum physics that is compatible with Maxwell's equations, with which we anticipate unified electromagnetic responses.
We find a large orbital diamagnetism only along the interplane direction and a nearly temperature-independent electrical conductivity of the order of $e^{2}/h$ per plane for the new 2D Dirac organic conductor, $\alpha$-(BETS)$_{2}$I$_{3}$, where BETS is bis(ethylenedithio)tetraselenafulvalene. 
Unlike conventional electrons in solids whose nonrelativistic effects bifurcate electric and magnetic responses, the observed orbital diamagnetism scales with the electrical conductivity in a wide temperature range. This demonstrates that an electromagnetic duality that is valid only within the relativistic framework is revived in solids.
\end{abstract} 
\maketitle
\newpage

Dirac electron systems (DESs) such as bismuth and graphene can be described by the Dirac equation and provide a platform to realize physical properties rooted in relativistic quantum physics~\cite{Wolff1964,Novoselov2005}. One prominent property of DESs is their large orbital diamagnetism, which reaches a maximum when the chemical potential is in the mass gap, unlike Landau diamagnetism in metals. This orbital diamagnetism, which is theoretically argued to originate from the interband effect of magnetic fields, is observed in three-dimensional (3D) DESs including bismuth and antiperovskites~\cite{Shoenberg1936,Wehrli1968,Fukuyama1970,Suetsugu2021}. This mechanism also applies to two-dimensional (2D) systems. The diamagnetism was observed for mass-produced graphene flakes~\cite{Li2015}. Here the random orientation of the flakes prevented separating the orbital diamagnetism agreeable with theory. Organic conductors have recently been found to realize 2D DESs with a bulk form such as $\alpha$-(BEDT-TTF)$_{2}$I$_{3}$ (BEDT-TTF = bis(ethylene)dithiotetrathiafulvalene)~\cite{Tajima2000,Katayama2006,Hirata2016}; however, this is realized only under high pressure, limiting magnetic experiments and making it difficult to obtain the absolute value of the susceptibility using SQUID magnetometers. 

The electric responses of 3D and 2D DESs show a sharp contrast. The uniform permittivity of bismuth is enhanced in accordance with its orbital diamagnetism~\cite{Edelman1975,Boyle1960}. On the other hand, graphene has no enhancement in the permittivity, but rather shows exotic quantized optical conductance and minimum dc conductivity through Klein tunneling~\cite{Novoselov2005,Katsnelson2006,Li2008}. The organic conductor $\alpha$-(BEDT-TTF)$_{2}$I$_{3}$ also shows temperature-independent conductivity on the order of $e^{2}/h$ per sheet~\cite{Tajima2007}.

These magnetic and electric responses of DESs can be viewed as parallel to quantum electrodynamics (QED), a relativistic quantum field theory, in which two responses are unified due to the Lorentz covariance (space--time symmetry). Indeed, for 3D DESs, the large orbital diamagnetism and the enhanced permittivity can be explained by charge renormalization in a unified way, demonstrating an electromagnetic duality specified by the space--time symmetry of the Dirac equation~\cite{Maebashi2017}. In contrast to 3D DESs, permittivity enhancement due to charge renormalization is absent in 2D DESs~\cite{Gonzalez1994}, although they do exhibit a quantized conductance. The dependence of the charge renormalization on the dimensionality of the system raises the fundamental question of the existence and nature of the universal phenomena in DESs irrespective of the system dimension. Therefore, the two principal goals of the study of 2D DESs are to determine the behavior of the orbital diamagnetism and its relationship with quantized electric responses, and to clarify whether the two responses can be described by a unified theory. Observation of the orbital diamagnetism would resolve these questions, and in order to obtain absolute values of the magnetic susceptibility, a bulk-form single crystal at ambient pressure would be ideal.

\begin{figure*}[hbt]
\includegraphics*[width=16cm]{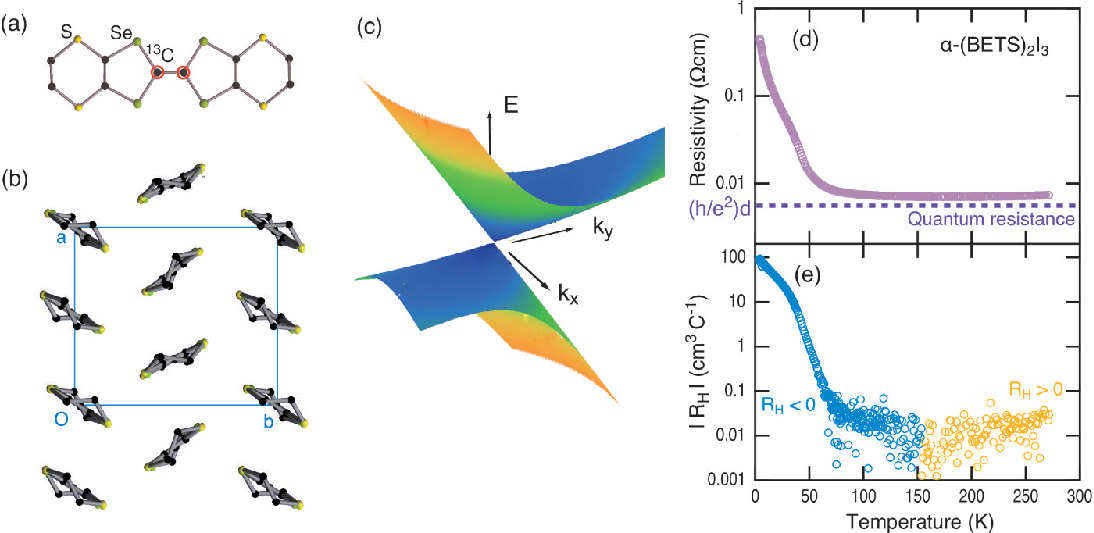}
\caption{(a) Molecular structure of bis(ethylenedithio)tetraselenafulvalene (BETS). Carbon atoms encircled in red are labeled by $^{13}$C for NMR experiments. (b) Molecular arrangement of conducting plane of $\alpha$-(BETS)$_{2}$I$_{3}$.
(c) Dirac cone band dispersions calculated by $ab$ $initio$ method. (d) Resistivity at ambient pressure. 
The bulk resistivity corresponding to the quantum sheet resistance, $(h/e^{2})d$, is shown as the dashed line. (e) Absolute values of the Hall coefficient.
}
\label{fig:Dirac}
\end{figure*}

In this Letter, we demonstrate the magnetic and transport properties of a newly identified 2D DES organic conductor with a bulk form at ambient pressure, $\alpha$-(BETS)$_{2}$I$_{3}$ with a strongly anisotropic magnetic susceptibility $\chi$. We discriminated a large orbital diamagnetism ($\chi_\mathrm{orb}$) from a spin susceptibility ($\chi_\mathrm{spin}$) by changing the field direction. $\chi_\mathrm{orb}$ shows quantitative agreement with the theory for $T>50$ K, where the dc-conductivity ($\sigma_\mathrm{dc}$) per sheet is independent of temperature with the value of $e^{2}/h$. The $-T\chi_\mathrm{orb}$ scales with $\sigma_\mathrm{dc}$ in a wide temperature range, showing an electromagnetic duality specified by the space--time symmetry in DESs, corresponding to the Lorentz covariance in  QED.

$\alpha$-(BETS)$_{2}$I$_{3}$ is composed of bis(ethylenedithio)tetraselenafulvalene (BETS) molecules that contain Se atoms (Fig.~\ref{fig:Dirac}~(a))~\cite{Inokuchi1995,Hiraki2011}. The structure is isomorphous with $\alpha$-(BEDT-TTF)$_{2}$I$_{3}$, as shown in Fig.~\ref{fig:Dirac}~(b). The molecular orbital of BETS is spatially larger than that of BEDT-TTF, which yields uncorrelated electron characteristics and prevents the instabilities toward charge ordering or excitonic orders observed in $\alpha$-(BEDT-TTF)$_{2}$I$_{3}$~\cite{Hirata2017}. This noninteracting character enables us to extract the ideal physics of DES through theoretical and quantitative analysis.

The full relativistic first-principles calculation provides a Dirac-like linear dispersion with a mass gap of $\approx 2$ meV (Fig.~\ref{fig:Dirac}~(c)) and an effective ``speed of light" of $ v\approx 5 \times 10^{4}$ m/s~\cite{Supplement,Kitou2021,Perdew1996,Wimmer1981}. The resistivity above 50 K is nearly independent of temperature, as has also been observed in the high-pressure massless Dirac phase of $\alpha$-(BEDT-TTF)$_{2}$I$_{3}$ (Fig.~\ref{fig:Dirac}~(d)). Here, Dirac electrons compensate for the temperature dependence of the mobility and the density of states~\cite{Tajima2000}, resulting in the temperature-independent resistivity corresponding to quantum sheet resistance, $(h/e^{2})d=4.6$ m$\Omega$ cm, where $d=17.8$ \AA \ is the interplane distance using the value of the lattice constant along $c$.

The resistivity increases upon cooling below 50 K without a phase transition, consistent with the mass gap but does not follow an activated temperature dependence. The Hall coefficient ($R_\mathrm{H}$) is small above 50 K, and its sign changes at $T= 150$ K from high-temperature positive (hole-like) values to low-temperature negative (electron-like) values, as shown in Fig.~\ref{fig:Dirac} (e). This indicates that the Fermi energy is in the mass gap but shifts slightly with temperature. At 150 K, the Fermi energy will be exactly at the midpoint of the gap.

\begin{figure*}[t]
\includegraphics*[width=16cm]{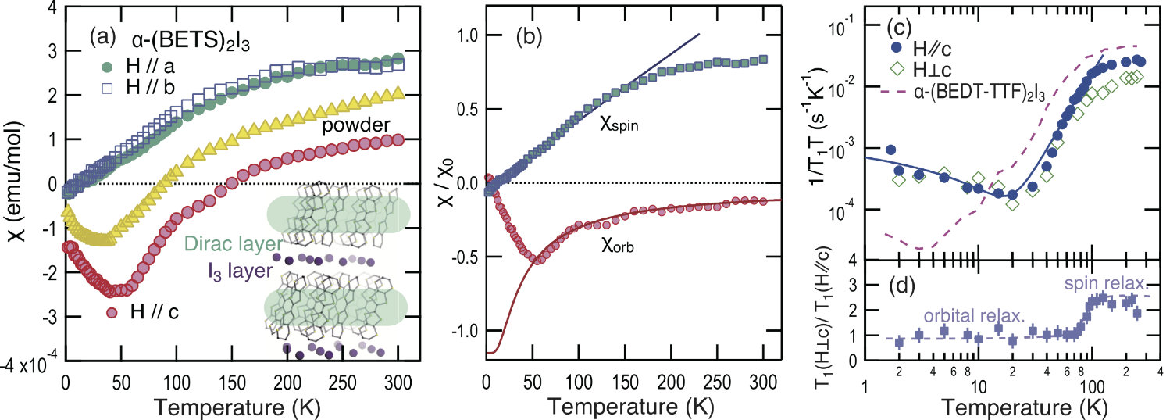}
\caption{Magnetic properties of $\alpha$-(BETS)$_{2}$I$_{3}$. (a) Magnetic susceptibilities ($\chi$) of a single-crystal measured for $H \parallel$ $a$, $b$, and $c$ directions and polycrystalline samples. In the inset, the hatched area is the $ab$-plane hosting the 2D Dirac electrons isolated by anion (I$_{3}$) layers. (b) Spin ($\chi_\mathrm{spin}$) and orbital ($\chi_\mathrm{orb}$) susceptibilities scaled by $\chi_{0}=\mu_{0}e^{2}/2\pi m_{0}d$. Solid curves are calculated susceptibilities for $\Delta=50 $ K using Eqs.~\eqref{eq: spin} and \eqref{eq: full temperature}. (c) Nuclear spin-lattice relaxation rate divided by temperature, $1/T_{1}T$, of $^{13}$C NMR. The solid curve is a functional form of $1/T_{1}T= aT^{2.5}+b\log(T^{*}/T)$. The dashed curve of $1/T_{1}T$ for an organic DES, $\alpha$-(BEDT-TTF)$_{2}$I$_{3}$, under high pressure is from Ref.~\cite{Hirata2017}. (d) Ratios of $1/T_{1}$'s for $H\parallel c$ and $H \perp c$ calculated using data in (c).
 }
\label{fig:mag}
\end{figure*}

We show in Fig.~\ref{fig:mag}~(a) the magnetic susceptibilities $\chi_{a,b,c}$ for $H\parallel a, b,$ and $c$ (interplane direction). The susceptibility $\chi_{a}$, which nearly agrees with $\chi_{b}$, decreases linearly upon cooling below 150 K, consistent with Dirac-type linear dispersion. A possible in-plane anisotropy of $\chi$ originating from the tilting of the Dirac cone is negligible; therefore, we can consider that $\chi_{a,b}$ solely depends on the density of states and the electronic correlation is negligible. A detailed formula for the spin contribution to $\chi$ is given below using $\chi_{0}$, the spin susceptibility of 2D nonrelativistic electron gas with the interplane distance $d$.
\begin{align}
\chi_\mathrm{spin}(T) 
= 
\frac{m^*}{m_0} \chi_0 \left(
\frac{2}{\beta\Delta} \ln \left( 2 \cosh \frac{\beta \Delta}{2} \right) 
- \tanh \frac{\beta \Delta}{2}
\right) , \label{eq: spin}
\end{align}
where $m_0$ is the electron mass, $\Delta=m^{*}v^{2}$ is the mass gap, and $\beta=1/k_{B}T$. $\chi_{0}=\mu_0e^{2}/2 \pi m_{0} d=1.57 \times 10^{{-7}}$ in SI units, where $\mu_0$ is the vacuum permeability. The chemical potential $\mu$ is set to be zero for simplicity (see Ref.~\cite{Supplement} for general $\mu$). In Fig.~\ref{fig:mag}~(b), we plot $\chi_\mathrm{spin}$ using $\Delta=50$ K, which provides $m^{*}=\Delta/v^2 =0.3m_{0}$ with the \textit{ab~initio} value of $v$, as well as the experimental $\chi_\mathrm{spin}^{\mathrm{(exp)}}= (\chi_{a}+\chi_{b})/2 $, and find that this simple formula quantitatively reproduce the experiments.

The magnetic susceptibility perpendicular to the conducting plane, $\chi_{c}$, is strongly suppressed and shows negative values below 150 K, which indicates an orbital diamagnetism, $\chi^\mathrm{(exp)}_\mathrm{orb}=\chi_{c}-\chi_{\mathrm{spin}}^{\mathrm{(exp)}} $, that emerges only along the $c$-direction. This diamagnetic $\chi_\mathrm{orb}$ was theoretically predicted for 2D DESs~\cite{McClure1956,Fukuyama2007}. In contrast to spins, which are conserved, the orbital currents that generate $\chi_\mathrm{orb}$ are not conserved.  In general, the susceptibility of a non-conserved quantity has a contribution from high-energy bands, $\chi_\mathrm{orb}^\mathrm{(exp)}(\infty)$, which is independent of temperature and irrelevant to the Dirac band. We estimated $\chi_\mathrm{orb}^\mathrm{(exp)}(\infty)$ by fitting $\chi_\mathrm{orb}^\mathrm{(exp)}(T)$ for $T> 100$ K with $\chi^\mathrm{(exp)}_\mathrm{orb}(T)=\mathrm{const} \times 1/T+\chi_\mathrm{orb}^\mathrm{(exp)}(\infty)$ (see below Eq.~(\ref{eq: singularity})). We plot in Fig.~\ref{fig:mag} (b) the temperature dependent component, $\chi_\mathrm{orb}(T)=\chi^\mathrm{(exp)}_\mathrm{orb}(T) -\chi^\mathrm{(exp)}_\mathrm{orb}(\infty)$.

For free electrons, a detailed formula for $\chi_\mathrm{orb}(T)$ in the presence of a mass gap is given by~\cite{Koshino2011}
\begin{align}
\chi_\mathrm{orb}(T) = 
- \frac{2}{3}  \frac{m_0}{m^*} \chi_0 \tanh \frac{\beta \Delta}{2}  ,
\label{eq: full temperature}
\end{align}
where the chemical potential $\mu$ is set to be zero (see Ref.~\cite{Supplement} for general $\mu$).
In Fig.~\ref{fig:mag}~(b), we plot experimental and calculated $\chi_\mathrm{orb}$ obtained using Eq.~\eqref{eq: full temperature} with the same parameters as those for $\chi_\mathrm{spin}$. 
We find quantitative agreements between the experimental and theoretical values as well as those for $\chi_\mathrm{spin}$ in the wide temperature range of $T>50$ K. Equation~\eqref{eq: full temperature} shows a crossover at $T\approx \Delta/k_{B}$ and is approximated as
\begin{align}
\chi_\mathrm{orb}(T) = 
- \frac{2}{3} \frac{m_0 v^2}{\mathrm{max} (\Delta, 2 k_{\rm B}T)}\, \chi_0, 
\label{eq: singularity}
\end{align}
so that $\chi_\mathrm{orb}T$ = const is expected for $T\gtrsim \Delta/k_\mathrm{B}$. 

The observed uncorrelated character of the 2D DES for $T> 50$ K and the deviation of $\chi_\mathrm{orb}(T)$ from Eq.~\eqref{eq: full temperature} are microscopically supported by $^{13}$C NMR. High-temperature Korringa-like $1/T_{1}T$ for $T>200$ K is significantly reduced following $1/T_{1}T\propto T^{\gamma}$ with $\gamma \approx2$ for $30 <T <100$ K as shown in Fig.~\ref{fig:mag} (c), which indicates a linear dispersion. Note that the observed $1/T_{1}T$ is five times smaller than that of $\alpha$-(BEDT-TTF)$_{2}$I$_{3}$, showing that the Dirac electrons in $\alpha$-(BETS)$_{2}$I$_{3}$ are relatively free of one-body renormalization of Coulomb repulsions~\cite{Hirata2017}. The increase in $1/T_{1}T$ below 20 K indicates other emergent relaxation mechanisms. Fig.~\ref{fig:mag}~(d) depicts the anisotropies of $1/T_{1}$, which we expect to be temperature-independent when the spin contribution $(1/T_{1})_\mathrm{spin}$ dominates $1/T_{1}$. The reduction of the anisotropy below 100 K, the onset temperature of the DES, coincides with that of $(1/T_{1})_\mathrm{spin}$. Since $^{13}$C does not couple with the electric field gradient, the most plausible source of the relaxation at low temperatures is the fluctuation of the orbital currents, which contributes to $1/T_{1}$ as $1/T_{1}=(1/T_{1})_\mathrm{spin}+(1/T_{1})_\mathrm{orb}$. Recent theories point out that $(1/T_{1})_\mathrm{orb}$ dominates $1/T_{1}$ in 3D Weyl materials~\cite{Dora2009,Maebashi2019}, but predict $(1/T_{1}T)_\mathrm{orb}\propto T$ for clean 2D DESs, which does not reproduce the experiments below 20 K~\cite{Maebashi2019}. Later, we will discuss a potential mechanism for the deviation related to $\chi_\mathrm{orb}$.

The correspondence between DES and QED relates Eq.~\eqref{eq: singularity} to the exotic quantized electric property in 2D DESs. In parallel to the Lorentz covariance in QED, we can show a duality between electric and magnetic responses in DES~\cite{Supplement,Peskin1995,Vignale2005}. For $|\mu| \leq \Delta$ and $T=0$, the static magnetic susceptibility $\chi_\mathrm{orb}$ is given exactly as
\begin{align}
\chi_\mathrm{orb} = - \frac{2}{\pi} \left(\frac{v}{c}\right)^2 \int_{2\Delta/\hbar}^{\infty} \frac{\sigma(\omega)}{\varepsilon_{0} \omega^2} d \omega,
\label{eq: duality} 
\end{align}
where $c$ and $\varepsilon_0$ are  the speed of light and vacuum permittivity, respectively~\cite{Supplement}. Here, $\sigma(\omega)$ is the dynamical electrical conductivity, which originates only from interband electron--hole excitations. Thus, this duality relation indicates that dynamical vacuum fluctuations (the creation and annihilation of virtual electron--hole pairs), or the interband effect across the mass gap, necessarily generate the orbital diamagnetism $\chi_\mathrm{orb} < 0$.

A dimensional analysis gives $\sigma (\omega) \propto (e^2/h) (\omega/v)^{D-2}$ for the massless limit of $\Delta \to 0$ in the $D$ dimensions. In three dimensions ($D=3$), Eq.~\eqref{eq: duality} leads to a logarithmic divergence in $\chi_\mathrm{orb}$ for $\Delta \to 0$, which corresponds to the well-known ultraviolet divergence in the charge renormalization of QED~\cite{Maebashi2017}. In two dimensions ($D=2$), on the other hand, there is no charge renormalization~\cite{Gonzalez1994}. The large diamagnetism $\chi_\mathrm{orb} \propto - 1/\Delta$ in Eq.~\eqref{eq: singularity} is therefore free from charge renormalization but closely linked to the $\omega$-independent electrical conductivity for $\Delta \to 0$, where it takes a universal value of  $\sigma_0=e^2/4 \hbar d$ (quantized optical conductance)~\cite{Gusynin2006,Li2008}. (In Table~\ref{tab:duality}, we summarize $\chi_\mathrm{orb}$ and $\sigma(\omega)$ as well as the permittivity $\varepsilon$ for 3D and 2D DESs.)
More precisely, using a detailed formula for $\sigma(\omega)$~\cite{Gusynin2006}, we find that the duality relation, Eq.~\eqref{eq: duality}, expresses $\chi_\mathrm{orb}$ in terms of the universal constant $\sigma_0$ as
\begin{align}
\chi_\mathrm{orb} =-\frac{4}{3 \pi} \left(\frac{v}{c}\right)^2 \!\! \frac{\hbar}{\varepsilon_0 \Delta} \, \sigma_0.
\label{eq: orbital} 
\end{align} 

\begin{table}[hbt]
\caption{\label{tab:duality}Electromagnetic responses of 3D and 2D DESs, which capture the salient nature of QED, i.e., the Lorentz covariance (space--time symmetry) and charge renormalization. An electromagnetic duality resulting from the space--time symmetry relates $\chi_\mathrm{orb}$ to $\sigma(\omega)$ through Eq.~\eqref{eq: duality}. The permittivity $\varepsilon(q, \omega)$ is renormalized at $q=\omega=0$ as $\varepsilon(0,0)=Z_{3}^{-1}\varepsilon_{0}$, where $Z_{3}$ is the charge renormalization factor. The enhancement of $-\chi_\mathrm{orb}$ originates from the enhanced $\varepsilon(0,0)$ for 3D DESs, whereas that for 2D DESs takes place with $Z_{3}=1$. For finite temperatures, $\chi_\mathrm{orb}$ and $\varepsilon(0, 0)$ are given by replacing $\Delta$ by $T$. Details are given in Ref.~\cite{Supplement}}
\begin{ruledtabular}
\begin{tabular}{rccc}
& $ \chi_\mathrm{orb}$ & $\sigma(\omega \gg 2\Delta/\hbar )$ & $\varepsilon (0, 0)/\varepsilon_{0}=Z_{3}^{-1}$\\ \hline
3D & $\propto \ln \Delta$ & $\approx (e^{2}/h)\omega/v$ & $\propto -\ln \Delta$\\
2D & $\propto -1/\Delta$ & $\approx e^{2}/hd$ & 1\\
\end{tabular}
\end{ruledtabular}
\end{table}

It is noteworthy that the conductivity unit $\sigma_{0}$ can be rewritten using the susceptibility unit $\chi_{0}$ as $\displaystyle \sigma_{0}=(\pi^{2} / Z_{0} \lambda_{e}) \chi_{0}$, where $Z_{0} = \sqrt{\mu_0/\varepsilon_0} \approx 120 \pi $ $\Omega $ is the impedance of free space and $\lambda_{e} = h/m_0 c$ is the Compton wavelength, leading to the equivalence of Eqs.~\eqref{eq: singularity} and~\eqref{eq: orbital}. This equivalence shows that $\chi_\mathrm{orb}$ scales with the universal electric conductance $\sigma_{0}d\approx e^{2}/h$ even for finite temperatures. 

The dc conductivity $\sigma_\mathrm{dc}\equiv \alpha \sigma_{0}$ ($\alpha$ is of the order of $1$) is difficult to determine theoretically, depending on the characteristics of the disorder~\cite{Shon1998,Adam2007,Ostrovsky2006,Ostrovsky2007}. $\alpha$ is naively given as $\alpha=8/\pi^{2}$~\cite{Ando2002} for 2D massless Dirac electrons but remains under debate for $T\neq 0$. The experimentally obtained values of $\sigma_\mathrm{dc}$'s for organic DES, in contrast, are independent of temperature both for $\alpha$-(BETS)$_{2}$I$_{3}$ and $\alpha$-(BEDT-TTF)$_{2}$I$_{3}$; the $\sigma_\mathrm{dc}$ values are approximately equal to $\sigma_\mathrm{dc}=14$ k$\Omega^{-1}$m$^{-1}$, corresponding to $\alpha\approx 4/\pi^{2}$~\cite{Tajima2007}.

\begin{figure}[h]
\includegraphics[width=6cm]{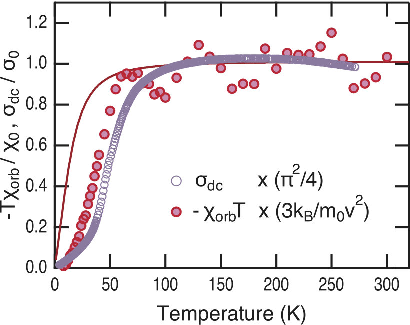}
\caption{Electromagnetic duality of $\alpha$-(BETS)$_{2}$I$_{3}$, where $-\chi_\mathrm{orb}T$ is scaled by $\sigma_\mathrm{dc}$ for a wide temperature range. The scaling factor is based on Eq.~\eqref{eq: singularity}. The solid curve is the calculated $-\chi_\mathrm{orb}T$.}
\label{fig:scale}
\end{figure}

We plot $-\chi_\mathrm{orb}T$ and $\sigma_\mathrm{dc}$, normalized by $\chi_{0}$ and $\sigma_{0}$, respectively, in Fig.~\ref{fig:scale}, and find that these electromagnetic responses are scaled in a wide temperature range, as anticipated from Eq.~\eqref{eq: orbital}. The observed electromagnetic duality manifests itself in the correspondence with the Lorentz covariance in QED. The interband effect across the mass gap in the presence of electromagnetic fields characterizes the physical properties.

We now discuss potential sources for the deviation of $\chi_\mathrm{orb}(T)$ from Eq.~\eqref{eq: full temperature}, although Eq.~\eqref{eq: spin} does reproduce $\chi_\mathrm{spin}(T)$ below 40 K. The most plausible source is a disorder in real materials, which raises a new problem related to the interaction of disorder and orbital currents in DESs. We found that the function $1/T_{1}T=a T^{2.5}+b \log (T^{*}/T)$ fits the $1/T_{1}T$, as shown in Fig.~\ref{fig:mag}~(c). The logarithmic increase upon cooling below 20 K does not originate from the electronic correlation, which enhances $1/T_{1}T$ for the whole temperature range, and the crossing of the $1/T_{1}T$ curves of $\alpha$-(BETS)$_{2}$I$_{3}$ and $\alpha$-(BEDT-TTF)$_{2}$I$_{3}$ at $T \approx 10$ K suggests disorder effects on $(1/T_{1})_\mathrm{orb}$. A similar moderate increase of $1/T_{1}T$ is observed for a 3D Weyl system~\cite{Yasuoka2017}, which has been theoretically analyzed considering the effects of impurities or temperature-dependent chemical potential to $(1/T_{1})_\mathrm{orb}$~\cite{Hirosawa2020,Okvatovity2019}. Likewise, an observed increase in $1/T_{1}T$ for a noninteracting 3D DES with $\Delta \approx 15$meV, Bi$_{0.9}$Sb$_{0.1}$~\cite{MacFarlane2014}, is closely related to our observation of the moderate increase in $1/T_{1}T$ below 20 K.
A related phenomenon is also observed for the transport properties: namely, unconventional negative magnetoresistance with a field dependence of the form $1-\rho(B)/\rho_{0}\propto -\sqrt{B}$~\cite{Supplement}. These deviations from an ideal 2D DES are observed solely for the orbital-related properties at low temperatures, which suggests a new problem in disordered orbital physics in DESs. Surprisingly, despite the deviation of $\chi_\mathrm{orb}$ from Eq.~\eqref{eq: full temperature} below 50 K, $\sigma_\mathrm{dc}$ approximately scales with $-\chi_\mathrm{orb}T$, thus maintaining the electromagnetic duality even at low temperatures where the effect of a disorder becomes crucial as shown in Fig.~\ref{fig:scale}. This suggests a possible relationship between the effects of disorder on $\chi_\mathrm{orb}$ and $\sigma_\mathrm{dc}$, which results in a less disturbed electromagnetic duality.


In summary, we identified the organic conductor, $\alpha$-(BETS)$_{2}$I$_{3}$, as a 2D DES at ambient pressure through electric and magnetic measurements of $\sigma_\mathrm{dc}$, $R_\mathrm{H}$, $\chi_\mathrm{spin}$, and $1/T_{1}$ of $^{13}$C NMR. The latter two magnetic responses show negligible electronic correlation, enabling us to study an ideal characteristics of DES. We found orbital diamagnetism ($\chi_\mathrm{orb}$) only along the interplane direction. We demonstrate that the equation $T\chi_\mathrm{orb}=\mathrm{const}$ holds approximately for $T>50$ K and $\chi_\mathrm{spin}\propto T$, and that small shifts from the gapless DES are well reproduced by the theory using a unique parameter, $\Delta=m^{*}v^{2}$, the mass gap for the DES. We found a unified electromagnetic responses in which $-T\chi_\mathrm{orb}$ scales with $\sigma_\mathrm{dc}\approx e^{2}/hd$ in a wide temperature range, as shown in Fig.~\ref{fig:scale}, consistent with an electromagnetic duality that is valid only within the relativistic framework.

\begin{acknowledgments}
We are grateful to H.~Fukuyama, H.~Sawa, T.~Morinari, Y.~Fuseya, H. Matsuura, I. Tateishi, and S. Ozaki for fruitful discussions. This work was supported by Grants-in-Aid for Scientific Research (20K03870, 21K03426, 18K03482, 18H01162, 19K21860, and 16H06346) from JSPS.
\end{acknowledgments}
%
  
\end{document}


\title{Supplemental Material:\\
``Large Diamagnetism and Electromagnetic Duality in Two-dimensional Dirac Electron System''}
\author{S.~Fujiyama}
\affiliation{RIKEN, Condensed Molecular Materials Laboratory, Wako 351-0198, Japan}
\author{H.~Maebashi}
\affiliation{Department of Physics, University of Tokyo, Tokyo 113-0033, Japan}
\author{N.~Tajima}
\affiliation{Department of Physics, Toho University, Funabashi 274-8510, Japan}
\author{T.~Tsumuraya}
\affiliation{POIE, Kumamoto University, Kumamoto 860-8555, Japan}
\author{H-B.~Cui}
\affiliation{RIKEN, Condensed Molecular Materials Laboratory, Wako 351-0198, Japan}
\author{M.~Ogata}
\affiliation{Department of Physics, University of Tokyo, Tokyo 113-0033, Japan}
\affiliation{Trans-scale Quantum Science Institute, University of Tokyo, Bunkyo-ku, Tokyo 113-0033, Japan}
\author{R.\ Kato}
\affiliation{RIKEN, Condensed Molecular Materials Laboratory, Wako 351-0198, Japan}
\date{5 May 2021}
\maketitle
\beginsupplement

\subsection{1. Experimental procedure and calculation of the band structure}

We obtained black plates of $\alpha$-(BETS)$_{2}$I$_{3}$ by galvanostatic electrochemical oxidation (with a constant current of 1.5 $\mu$A) of BETS in chlorobenzene and 2\%(vol) methanol solution of $^{13}$C$_{2}$-BETS and (n-C$_{4}$H$_{9}$)NI$_{3}$ using Pt electrodes ($\phi$1 mm) at 30 $^{\circ}$C. For $^{13}$C-NMR measurements, we synthesized $^{13}$C-enriched ($>99.9$ \%) BETS molecules, where the central double-bonded carbon atoms were substituted by $^{13}$C isotopes using $^{13}$C-enriched triphosgene as a starting compound.  Resistance was measured by a conventional dc method with six probes applying an electrical current along $a$. For the Hall conductivity measurements, the magnetic fields were applied in the direction normal to the 2D plane. Susceptibility was measured using a SQUID magnetometer. NMR experiments were performed under magnetic fields of $\approx 8$ T. 

\begin{figure*}[hbt]
  \includegraphics*[width=16cm]{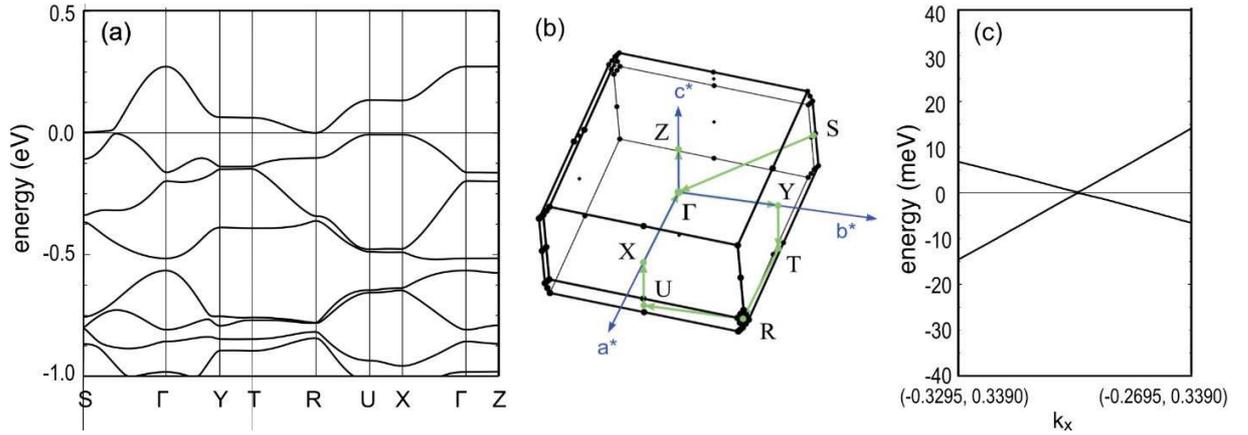}
\caption{(a) Band dispersion of $\alpha$-(BETS)$_{2}$I$_{3}$. (b) Schematic plot of the Brillouin zone. The dispersion in (a) is obtained along the green line connecting the symmetric points in the first Brillouin zone, which are given by $\Gamma=(0, 0, 0)$, $\mathrm{X} = (0.5, 0, 0)$, $\mathrm{Y}=(0, 0.5, 0)$, $\mathrm{Z}=(0, 0, 0.5)$, $\mathrm{S}= (-0.5, 0.5, 0)$, $\mathrm{T}=(0, 0.5, -0.5)$, $\mathrm{R}= (0.5, 0.5, -0.5)$, $\mathrm{U}=(0.5, 0, -0.5)$ in unit  of $(2\pi/\mathbf{a}, 2\pi/\mathbf{b}, 2\pi/\mathbf{c})$. The lattice parameters were obtained by x-ray diffraction as $a=9.1, b=10.7$, $c=17.4$ \AA, $\alpha=96.7$, $\beta=98.0$, and $\gamma=90.9$ \textit{deg.}~\cite{Kitou2021}
(c) Dispersion along the lines $(k_x, 0.3390,0)$ in the reciprocal space. This intersection creates a discrete contact point known as a Dirac point at $\vb*{k}_D$ = ($k_x$, $k_y$) = ($\mp$0.2958, $\pm$0.3390), which is located at the $E_F$.}
\label{fig:band}
\end{figure*}

We performed a band structure calculation for $\alpha$-(BETS)$_{2}$I$_{3}$ using the first-principles density-functional theory (DFT) method. The crystal structure is a triclinic structure with the space group of \textit{P}\={1}, which was measured at 30 K with synchrotron x ray diffraction~\cite{Kitou2021}. The Kohn--Sham equations are self-consistently solved using the all-electron full-potential linearized augmented plane wave (FLAPW) method implemented in the QMD-FLAPW code~\cite{Wimmer1981}. 
The exchange-correlation functional used is the generalized gradient approximation (GGA) proposed by Perdew, Burke, and Ernzerhof (PBE)~\cite{Perdew1996}. In Fig.~\ref{fig:band}, we plot the band structure in the absence of spin-orbit coupling (SOC), by which a gapless Dirac point is obtained. The dimensions of the $\vb*{k}$-point mesh used were $6\times  6 \times 2$ for the self-consistent loop. We used a high-dense $\vb*{k}$-point mesh for plotting the 3D band structures shown in Fig.~1(c). The cutoff energies for LAPW basis and the potential and density was 282 Ry. We set the muffin-tin (MT) sphere radii as 1.26, 0.75, 2.00, and 2.27 Bohr for C, H, S, and Se atoms, respectively. The electronic states up to C $(2s)^{2}$, S $(2p)^{6}$, Se $(3p)^{6}$, and I $(4d)^{10}$ were treated as core electrons, which are predominantly confined to the MT spheres. 

We estimated the velocity, $v$, by fitting the areas of isoenergy conic sections of the Dirac cone ($S$) in the $\vb*{k}$ space by the following procedure. We used the first-principles calculation without SOC, which has a Dirac crossing point. The dispersion obtained with SOC approaches asymptotically to that without SOC above the mass gap.

\begin{figure*}[htb]
   \includegraphics*[width=12cm]{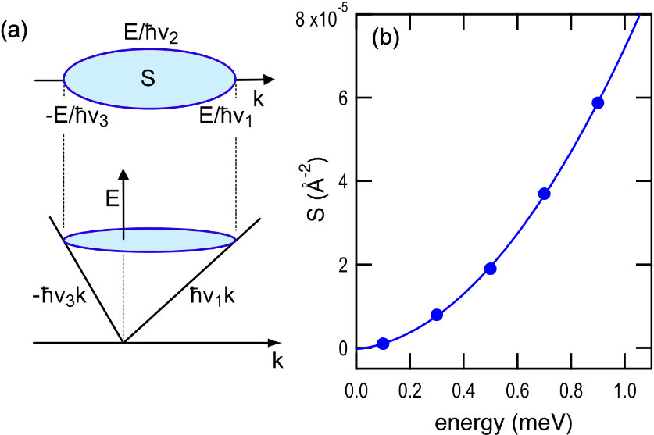}
  \caption{(a) Schematic view of the relation between $S$($E$) and the velocities. (b) Energy dependent areas of the conic sections of the Dirac cone obtained by the band calculation. }
  \label{fig:velocity}
  \end{figure*}

We define $v_i$ $(i=1, 2, 3)$ as the velocities along (perpendicular to) the tilting direction as $i=1$ ($i=2$), and $i=3$ along the opposite direction to $v_{1}$ as shown in Fig.~\ref{fig:velocity}~(a). The isoenergy conic sections are given as $\displaystyle k_{i}=\frac{E}{\hbar v_{i}}$, which gives $S$ as,
\begin{align}
  S=\frac{\pi E^{2}}{2\hbar^{2}}\frac{1}{v_{2}} \qty(\frac{1}{v_{1}}+\frac{1}{v_{3}}).
  \label{eq: tiltvelocity}
  \end{align}
We plot in Fig.~\ref{fig:velocity}~(b) the energy-dependent areas of the isoenergy conic sections and find $S \propto E^2$ relation.

In the absence of the tilting of the cone, $S$ is written using a single velocity $v$ as,
\begin{align}
  S=\frac{\pi E^{2}}{\hbar^2 v^2},
  \label{eq: velocity}
  \end{align}
and gives a velocity as $v=5.0\times 10^{4}$ m/s. This $v$ is an effective velocity representing the tilted cone, which works well to reproduce magnetic susceptibility that solely depends on the density of states. By comparing Eqs.~(\ref{eq: tiltvelocity}) and (\ref{eq: velocity}), $v$ has a relation with $v_{i}$ for a tilted Dirac cone as,
\begin{align}
  \frac{1}{v}=\sqrt{\frac{1}{2v_{2}}\qty( \frac{1}{v_{1}}+\frac{1}{v_{3}})}.
  \label{eq: velocityrelation}
  \end{align}

This procedure can be rewritten using a tilting parameter. We consider a massless tilted Dirac Hamiltonian, $\displaystyle  H=v_{0}(\eta p_{x}\sigma_{0}+p_{x}\sigma_{x}+p_{y}\sigma_{y})$, where $v_0$ is the velocity for an untilted Dirac cone, $\eta$ is a tilting parameter, and $\sigma_{0}$ is a unit matrix. The velocities can be written using $v_{0}$ and $\displaystyle \eta$ as $\displaystyle v_{1}=(1 + \eta )v_{0}$, $\displaystyle v_{3}=(1-\eta) v_{0}$, and $\displaystyle v_{2}=\sqrt{1-\eta^2}v_{0}.$
Using this expression the effective velocity, $v$, is given as,
\begin{align}
  v=v_{0}(1-\eta^2)^{-3/4}.
\end{align}
%

\subsection{2. Spin and orbital susceptibilities: Derivation of Eqs.~(1) and~(2)}

We consider 2D DES with an effective `speed of light' $v$ and a mass gap $\Delta = m^* v^2$ described by the Hamiltonian
\begin{align}
H =v(p_{x}\sigma_{x}+p_{y}\sigma_{y}) + \Delta \sigma_{z}, 
\label{eq: Hamiltonian}
\end{align}
where $p_{x} = - i \hbar \partial_x$, $p_{y} = - i \hbar \partial_y$, and $\sigma_x$, $\sigma_y$, and $\sigma_z$ are the Pauli matrices. We assume that the valley and spin degeneracies are given by $g_\mathrm{v}$ and $g_\mathrm{s}$, respectively, and the 2D planes are stacked in the perpendicular direction with an interplane distance $d$ ($g_\mathrm{v} = g_\mathrm{s} = 2$ and $d=17.8$ \AA \ for $\alpha$-(BETS)$_{2}$I$_{3}$). Because the Hamiltonian, Eq.~\eqref{eq: Hamiltonian}, describes a non-interacting system, the spin and orbital susceptibilities, which depend on temperature $T$ and the chemical potential $\mu$, can be given as
\begin{align}
\chi_\mathrm{spin}(T,\mu) &= \int_{-\infty}^{\infty} d \varepsilon 
\left(- \frac{\partial f (\varepsilon)}{\partial \varepsilon} \right)  \chi_\mathrm{spin}(0,\varepsilon),
\label{eq: noninteracting_spin}
\\
\chi_\mathrm{orb}(T,\mu) &= \int_{-\infty}^{\infty} d \varepsilon 
\left(- \frac{\partial f (\varepsilon)}{\partial \varepsilon} \right)  \chi_\mathrm{orb}(0,\varepsilon),
\label{eq: noninteracting_orb}
\end{align}
where $f (\varepsilon) = 1/(e^{\beta (\varepsilon - \mu)} + 1)$ is the Fermi distribution function with $\beta = 1/k_{\rm B}T$~\cite{Vignale2005}. The zero-temperature spin susceptibility $\chi_\mathrm{spin}(0,\varepsilon)$ is proportional to the density of states $N(\varepsilon)$ as $\chi_\mathrm{spin}(0,\varepsilon) = \mu_{\rm B}^2 N (\varepsilon)$ with the Bohr magneton $\mu_{\rm B} = e\hbar/2m_0$, so that 
\begin{align}
\chi_\mathrm{spin}(0,\varepsilon) = 
\frac{m^* |\varepsilon|}{m_0\, \Delta} \chi_0\,  \theta ( |\varepsilon| - \Delta ) ,
\label{eq: T=0_spin}
\end{align}
where $\chi_{0}=\mu_0e^{2}g_\mathrm{v}g_\mathrm{s}/8 \pi m_{0} d$ is the spin susceptibility of 2D nonrelativistic electron gas with the interplane distance $d$ in SI units ($m_0$ and $\mu_0$ are the electron mass and  vacuum permeability, respectively) and $\theta(x)$ is the Heaviside step function. On the other hand, the zero-temperature orbital susceptibility $\chi_\mathrm{orb}(0,\varepsilon)$ is given by
\begin{align}
\chi_\mathrm{orb}(0,\varepsilon) = 
- \frac{2}{3}\frac{m_0}{m^*} \chi_0 \, \theta ( \Delta - |\varepsilon| ) ,
\label{eq: T=0_orb}
\end{align}
where the factor $2/3$ is related to the fact that the Landau diamagnetism is equivalent to $-1/3$ times Pauli paramagnetism for the nonrelativistic electron gas~\cite{Koshino2011}. By substituting Eqs.~\eqref{eq: T=0_spin} and \eqref{eq: T=0_orb} into Eqs.~\eqref{eq: noninteracting_spin} and \eqref{eq: noninteracting_orb}, respectively, we obtain
\begin{align}
\chi_\mathrm{spin}(T, \mu) =& \frac{m^*}{m_0} \chi_0 \left(
\frac{1}{e^{\beta(\Delta - \mu)}+1} +  \frac{1}{e^{\beta(\Delta + \mu)}+1} 
\right.
\nonumber
\\
& \left. + \frac{1}{\beta\Delta} 
\left[ \ln \left(1+e^{-\beta(\Delta - \mu)}\right) 
+ \ln \left(1+e^{-\beta(\Delta + \mu)}\right) \right] \right), 
\label{eq: chi_spin}
\\
\chi_\mathrm{orb}(T, \mu) =& 
- \frac{2}{3} \frac{m_0}{m^*} \chi_0
\left(1 - \frac{1}{e^{\beta(\Delta - \mu)}+1} - \frac{1}{e^{\beta(\Delta + \mu)}+1}\right). 
\label{eq: chi_orb}
\end{align}
When $\mu = 0$, by using hyperbolic functions, we obtain
\begin{align}
\chi_\mathrm{spin}(T) 
&= \frac{m^*}{m_0} \chi_0 \left(
\frac{2}{\beta\Delta} \ln \left( 2 \cosh \frac{\beta \Delta}{2} \right) 
- \tanh \frac{\beta \Delta}{2} \right) , 
\tag{1}
\\
\chi_\mathrm{orb}(T) 
&=- \frac{2}{3}  \frac{m_0}{m^*} \chi_0 \tanh \frac{\beta \Delta}{2}.
\tag{2}
\end{align}

\subsection{3. Electromagnetic duality: Derivation of Eqs.~(4) and~(5)}

In a standard nonrelativistic field theory~\cite{Vignale2005}, the linear response of electric current ${\boldsymbol  J} ({\boldsymbol q},\omega)$ to the vector potential ${\boldsymbol  A} ({\boldsymbol q},\omega)$ 
for a $D$-dimensional isotropic system can be written as
\begin{align}
J_{i} ({\boldsymbol q},\omega) 
&= K_{ij} ({\boldsymbol q},\omega) A_{j} ({\boldsymbol q},\omega), 
\label{eq: current}
\end{align}
where $i,j = 1,2,\cdots,D$ and the repeated indexes are to be summed. The response function $K_{ij} ({\boldsymbol q},\omega)$ is separated into the longitudinal component $K_\mathrm{L} (q, \omega)$ and the transverse component $K_\mathrm{T} (q, \omega)$ as 
\begin{align}
K_{ij} ({\boldsymbol q}, \omega) = \frac{q_i q_j}{q^2} K_\mathrm{L} (q, \omega) 
+ \left( \delta_{ij} - \frac{q_i q_j}{q^2} \right) K_\mathrm{T} (q, \omega). 
\label{eq: response}
\end{align}
The longitudinal and transverse components describe the electric and magnetic properties, respectively. In particular, the dynamical electrical conductivity $\sigma(\omega)$ and orbital magnetic susceptibility $\chi_\mathrm{orb}$ are given by
\begin{align}
\sigma(\omega) &= \frac{\mathrm{Im} K_\mathrm{L} (0, \omega)}{\omega},
\label{eq: sigma_nonrelativistic}
\\
\chi_\mathrm{orb} &= \mu_0 \lim_{q\to 0} \frac{K_\mathrm{T} (q, 0)}{q^2}.
\label{eq: chi_nonrelativistic}
\end{align}
Since $K_\mathrm{L} (q, \omega)$ and $K_\mathrm{T} (q, \omega)$ are independent of each other in general, there exists no universal relationship between the electric and magnetic properties of nonrelativistic systems. 

In parallel to $(D\!+\!1)$-dimensional QED~\cite{Peskin1995}, on the other hand, we can obtain the linear response of a $(D\!+\!1)$-current $J^\mu = (v \rho, {\boldsymbol  J})$ to a $(D\!+\!1)$-potential $A_\mu = (\phi/v, -{\boldsymbol A})$ with the charge density $\rho$ and the scalar potential $\phi$ for $D$-dimensional DES as
\begin{align}
J^{\mu} ({\boldsymbol q},\omega) 
&= -\varepsilon_0 v^2 \Pi^{\mu\nu} ({\boldsymbol q},\omega) A_{\nu} ({\boldsymbol q},\omega), 
\label{eq: four-current}
\end{align}
where $\varepsilon_0$ is the vacuum permittivity and $\mu,\nu = 0,1,2,\cdots,D$ with the repeated indexes to be summed. The important point is that for the chemical potential inside the mass gap and zero temperature, 
the `vacuum' polarization tensor $\Pi^{\mu\nu} ({\boldsymbol q},\omega)$ takes the following `Lorentz covariant' form with $Q^\mu = (\omega/v,{\boldsymbol q})$ and $g^{\mu\nu} = \mathrm{diag} (1,-1, \cdots,-1)$~\cite{Maebashi2017}: 
\begin{align}
\Pi^{\mu\nu} ({\boldsymbol q},\omega) &= \left( Q^2 g^{\mu\nu} - Q^{\mu}Q^{\nu} \right) \Pi (Q^2),
\label{eq: Lorentz covariance}
\end{align}
where an analytic function $\Pi (Q^2)$ of $Q^2 =\omega^2/v^2 - q^2$ satisfies the Kramers--Kronig relation
\begin{align}
\mathrm{Re}\, \Pi\left( \frac{\omega^2}{v^2} - q^2 \right) 
&= - \frac{1}{\pi} \int_{2\Delta/\hbar}^\infty \frac{2 \omega'}{\omega^2 - \omega'^2} 
\mathrm{Im}\, \Pi\left( \frac{\omega^2}{v^2} - q^2 \right) d\omega' .
\label{eq: Kramers-Kronig} 
\end{align} 
By comparing Eqs.~\eqref{eq: current} and \eqref{eq: four-current}, we find that the response function $K_{ij}$ of DES can be described by the unique function $\Pi$ as
\begin{align}
K_{ij} ({\boldsymbol q}, \omega) = - \varepsilon_0 v^2
\left[ \left( \frac{\omega^2}{v^2} - q^2 \right) \delta_{ij} + q_i q_j \right] 
\Pi\left( \frac{\omega^2}{v^2} - q^2 \right) .
\end{align}
Both the longitudinal and transverse components in Eq.~\eqref{eq: response} are therefore related to this unique function as
\begin{align}
K_\mathrm{L} (q, \omega) &= -\varepsilon_0 \omega^2\Pi\left( \frac{\omega^2}{v^2} - q^2 \right) , 
\label{eq: longitudinal_Dirac}
\\
K_\mathrm{T} (q, \omega) &= -\varepsilon_0 \left( \omega^2 - v^2 q^2 \right) \Pi\left( \frac{\omega^2}{v^2} - q^2 \right) ,
\label{eq: transverse_Dirac}
\end{align}
leading to an exact relationship between them given by
\begin{align}
\left( \omega^2 - v^2 q^2 \right) K_\mathrm{L} (q, \omega) &= 
\omega^2 K_\mathrm{T} (q, \omega) .
\label{eq: duality_general}
\end{align}
Since $K_\mathrm{L} (q, \omega)$ and $K_\mathrm{T} (q, \omega)$ correspond to the electric and magnetic response functions, respectively, Eq.~\eqref{eq: duality_general} describes the electromagnetic duality in DES.

By substituting Eqs.~\eqref{eq: longitudinal_Dirac} and \eqref{eq: transverse_Dirac} into Eqs.~\eqref{eq: sigma_nonrelativistic} and \eqref{eq: chi_nonrelativistic}, respectively, we obtain
\begin{align}
\sigma(\omega) &= - \varepsilon_0 \omega \, \mathrm{Im}\, \Pi\left( \frac{\omega^2}{v^2} \right), 
\label{eq: sigma_relativistic}
\\
\chi_\mathrm{orb} &= \left( \frac{v}{c} \right)^2 \Pi\left( 0 \right) ,
\label{eq: chi_relativistic}
\end{align}
where $c= 1/\sqrt{\varepsilon_0 \mu_0}$ is the speed of light. Using the Kramers--Kronig relation, Eq.~\eqref{eq: Kramers-Kronig}, for Eqs.~\eqref{eq: sigma_relativistic} and \eqref{eq: chi_relativistic}, we obtain
\begin{align}
\chi_\mathrm{orb} = - \frac{2}{\pi} \left( \frac{v}{c} \right)^2 \int_{2\Delta/\hbar}^{\infty} \frac{\sigma(\omega)}{\varepsilon_0 \, \omega^2} \, d \omega .
\tag{4}
\label{Eq. 4}
\end{align}
Alternatively, by using Eqs.~\eqref{eq: Lorentz covariance} and~\eqref{eq: chi_relativistic}, we can relate $\chi_\mathrm{orb}$ to $\Pi^{00}({\boldsymbol q}, \omega) = -q^2\,  \Pi (\omega^2/v^2 - q^2)$, the density--density response function (multiplied by $e^2/\varepsilon_0$), as
\begin{align}
\chi_\mathrm{orb} = - \left( \frac{v}{c} \right)^2 \lim_{q \to 0} \frac{\Pi^{00}({\boldsymbol q}, 0)}{q^2} .
\label{eq: density-density}
\end{align}
These equations are equivalent to each other, where the gauge invariance guarantees  
$\sigma(\omega) = \varepsilon_0 \lim_{q \to 0} (\omega/q^2) \, \mathrm{Im}\,\Pi^{00}(q, \omega)$ 
and $\Pi^{00}(q, \omega)$ satisfies the Kramers--Kronig relation. They correspond to a special limiting case of Eq.~\eqref{eq: duality_general} and are available both in 2D and 3D DESs.

Note that the duality relation, Eq.~\eqref{Eq. 4}, can also be described in terms of the permittivity  $\varepsilon (q,\omega) = \varepsilon_0 [ 1 + V(q) \Pi^{00}({\boldsymbol q}, \omega)]$ with the Fourier transform $V(q)$ of the mutual Coulomb interaction (devided by $e^2/\varepsilon_0$), but then it depends on the dimension. This can be associated with the charge renormalization factor $Z_3$ that depends on the dimension as
\begin{align}
\frac{1}{Z_3} \equiv \frac{\varepsilon (0,0)}{\varepsilon_0} = \left\{ 
\begin{array}{ll}
{\rm const} \times \displaystyle{\ln (E_c/\Delta)} &\quad \mbox{(3D)}
\\
1 & \quad \mbox{(2D)}
\end{array}
\right.,
\label{eq: renormalization}
\end{align}
where $E_c$ is the ultraviolet (bandwidth) cutoff for QED (DES).
For $V(q) = q^{-2}$ in 3D DESs, $\varepsilon (q,\omega) / \varepsilon_0 = 1 - \Pi (\omega^2/v^2 - q^2)$ and Eq.~\eqref{eq: chi_relativistic} gives
\begin{align}
\chi_\mathrm{orb} = - \left( \frac{v}{c} \right)^2
\left( \frac{\varepsilon (0,0)}{\varepsilon_0} -1 \right) \quad \mbox{for 3D}. 
\end{align}
Thus, in 3D, $\chi_\mathrm{orb}$ has a large negative value in proportion to an enhancement in the permittivity for $\Delta \to 0$, corresponding to the well-known ultraviolet logarithmic divergence in charge renormalization of QED~\cite{Maebashi2017}. For $V(q) = (2q)^{-1}d$ in 2D, on the other hand, $\varepsilon (q,\omega) / \varepsilon_0 = 1 - (qd/2)\Pi (\omega^2/v^2 - q^2)$ and therefore $\varepsilon (0,\omega) / \varepsilon_0 = 1$, which leads to the absence of charge renormalization, $Z_3 = 1$, in Eq.~\eqref{eq: renormalization}. 
In this case,  Eq.~\eqref{eq: chi_relativistic} gives
\begin{align}
\chi_\mathrm{orb} = - \left( \frac{v}{c} \right)^2 \lim_{q \to 0} \frac{2}{q d}
\left( \frac{\varepsilon (q,0)}{\varepsilon_0} -1 \right) \quad \mbox{for 2D}. 
\label{eq: 2D epsilon}
\end{align}
Thus, in 2D, the large diamagnetism comes from the $q$ dependence of the static permittivity $\varepsilon(q,0)$, whereas the uniform permittivity is free from the interaction as $\varepsilon(0,\omega) = \varepsilon_0$.  

For the 2D DES described by the Hamiltonian, Eq.~\eqref{eq: Hamiltonian}, 
$\sigma(\omega)$ is given by  
\begin{align}
\sigma(\omega) = \sigma_0 \left( 1 + \frac{4 \Delta^2}{\hbar^2 \omega^2} \right) 
\theta (\hbar^2 \omega^2 - 4 \Delta^2 ) ,
\label{eq: conductivity}
\end{align} 
where $\sigma_0 = e^2g_\mathrm{v}g_\mathrm{s}/16\hbar d$~\cite{Gusynin2006}. 
Particularly for $\Delta \to 0$, $\sigma(\omega)$ becomes independent of $\omega$ and quantized as $\sigma(\omega) = \sigma_0$.
By substituting Eq.~\eqref{eq: conductivity} into Eq.~\eqref{Eq. 4}, we obtain
\begin{align}
\chi_\mathrm{orb} =-\frac{4}{3 \pi} \left(\frac{v}{c}\right)^2 \!\! \frac{ \hbar}{\varepsilon_0 \Delta} \, \sigma_0.
\tag{5}
\end{align} 
Since $\sigma_0$ is a universal constant,  $\chi_\mathrm{orb}$ takes a large negative value in proportion to $1/\Delta$ for $\Delta \to 0$.

\subsection{4. Magnetoresistance of $\alpha$-(BETS)$_{2}$I$_{3}$}
\begin{figure*}[ht]
\includegraphics*[width=13cm]{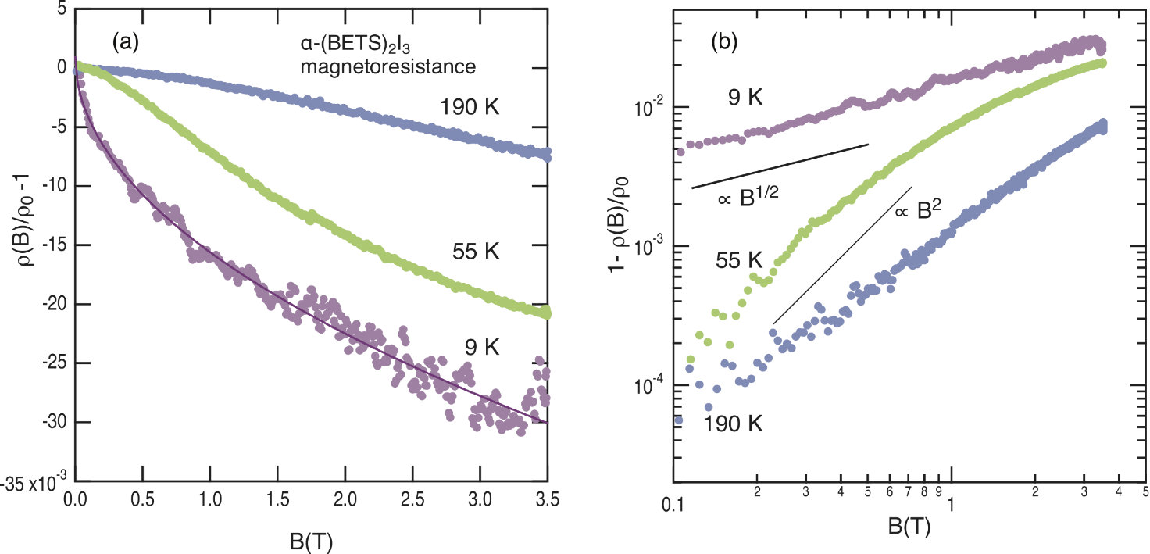}
\caption{Magnetoresistance of $\alpha$-(BETS)$_{2}$I$_{3}$}
\label{fig:MR}
\end{figure*}

We found negative magnetoresistance for a wide temperature range. We show in Fig.~\ref{fig:MR}~(a) the field-dependent resistivity $\rho(B)$ normalized by $\rho_{0}=\rho(B=0)$ for $T=9$, 55, and 190 K. The logarithmic plots of $\rho(B)$ (Fig.~\ref{fig:MR}~(b)) show that $\gamma$ in a function $1-\rho(B)/\rho_{0}\propto B^{\gamma}$ changes from $\gamma \approx 1.5$ at 190 K (temperature-independent $\rho$ region) to $\gamma \approx 0.5$ at 9 K (semiconducting region), suggestive of an effect of disorder whose theory remains room to investigate for 2D DESs. The function $1-\rho(B)/\rho_{0}\propto \sqrt{B}$ is the formula of Anderson-localization for three-dimensional systems. However, the DES keeps two-dimensionality because angular-dependent interlayer magnetoresistance estimates the transfer energy only as $t \approx 4$ K = 0.4 meV for $\alpha$-(BEDT-TTF)$_{2}$I$_{3}$, and a similar value is expected for $\alpha$-(BETS)$_{2}$I$_{3}$. At low temperatures, semiconducting in-plane transport due to disorder may cause a relative interplane correlation.

%